\documentclass[sigconf]{acmart}
\AtBeginDocument{%
  \providecommand\BibTeX{{%
    \normalfont B\kern-0.5em{\scshape i\kern-0.25em b}\kern-0.8em\TeX}}}

\copyrightyear{2022}
\acmYear{2022}
\setcopyright{rightsretained}
\acmConference[CHI '22]{CHI Conference on Human Factors in Computing
Systems}{April 29-May 5, 2022}{New Orleans, LA, USA}
\acmBooktitle{CHI Conference on Human Factors in Computing Systems
(CHI '22), April 29-May 5, 2022, New Orleans, LA, USA}
\acmDOI{10.1145/3491102.3517485}
\acmISBN{978-1-4503-9157-3/22/04}

\usepackage{enumitem}
\usepackage{stfloats}

\newcommand{\quot}[1]{``\emph{#1}''}

\makeatletter
\def\subsubsection{\@startsection{subsubsection}{3}%
  \z@{.5\linespacing\@plus.7\linespacing}{.1\linespacing}%
  {\normalfont\itshape}}
\makeatother

\begin{document}

\title{CrossData: Leveraging Text-Data Connections for \\ Authoring Data Documents}

%
\author{Chen Zhu-Tian}
\orcid{0000-0002-2313-0612}
\affiliation{%
  \institution{University of California San Diego}
  \city{La Jolla, CA}
  \country{USA}
}
\email{zhutian@ucsd.edu}

\author{Haijun Xia}
\orcid{0000-0002-9425-0881}
\affiliation{%
  \institution{University of California San Diego}
  \city{La Jolla, CA}
  \country{USA}
}
\email{haijunxia@ucsd.edu}


\begin{abstract}
Data documents play a central role in recording, presenting, and disseminating data. Despite the proliferation of applications and systems designed to support the analysis, visualization, and communication of data, writing data documents remains a laborious process, requiring a constant back-and-forth between data processing and writing tools. Interviews with eight professionals revealed that their workflows contained numerous tedious, repetitive, and error-prone operations. The key issue that we identified is the lack of persistent connection between text and data. Thus, we developed CrossData, a prototype that treats text-data connections as persistent, interactive, first-class objects. By automatically identifying, establishing, and leveraging text-data connections, CrossData enables rich interactions to assist in the authoring of data documents. An expert evaluation with eight users demonstrated the usefulness of CrossData, showing that it not only reduced the manual effort in writing data documents but also opened new possibilities to bridge the gap between data exploration and writing.
\end{abstract}





\begin{teaserfigure}
  \includegraphics[width=\textwidth]{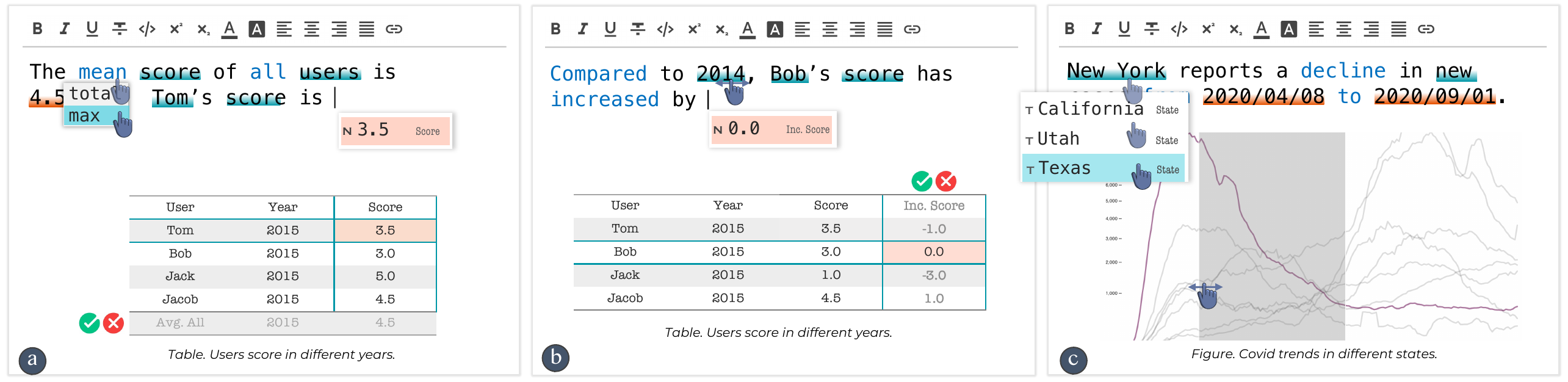}
  \caption{CrossData leverages text-data connections to enable users to efficiently retrieve (a), compute (b), interactively explore data (a, b, c), and adjust tables (a, b) and charts (c) during their writing processes, while also automatically maintaining data consistency between their text, data, tables, and charts.}
  \Description{CrossData leverages text-data connections to enable users to efficiently retrieve (a), compute (b), interactively explore data (a, b, c), and adjust tables (a, b) and charts (c) during their writing processes, while also automatically maintaining data consistency between their text, data, tables, and charts.}
  \label{fig:teaser}
\end{teaserfigure}

\maketitle

\section{Introduction}

Data documents employ text, tables, and visualizations to report findings from data analyses and present data-rich narratives, and are an indispensable component of every domain that uses data, such as scientific research, finance, public health, education, and journalism. As our world becomes increasingly data-driven, there has been a surge in the variety of data documents (e.g., data-rich documents~\cite{Badam2019}, data-driven articles~\cite{Sultanum2021}, and interactive articles~\cite{Conlen2018}), as well as in the research that has sought to support the authoring and consumption experiences of data documents. 

However, despite the proliferation of applications and systems that have been designed to support data analyses, visualization, and communication, authoring data documents remains a laborious task. During a typical workflow, a user will explore their data by performing data analysis operations (e.g., filtering, sorting, creating tables and charts, etc.) to generate insights using data processing tools and then they will synthesize the insights into a document using a word processing application. During this process, the user will need to switch back and forth between applications to take notes about the insights they discover, retrieve data from data processing tools and enter it into their document, as well as ensure that there is consistency between the data reported in their document and their underlying dataset. As the user's underlying data is updated or they iteratively refine, explore, and change their insights, the user will need to re-analyze their data, refine the corresponding tables and charts, and carefully identify and revise any out of date data in their document. This workflow is not only error-prone, but also requires significant manual and cognitive effort. 

The key reason that such tedious and ineffective workflows exist is due to the lack of persistent bindings or connections that exist between the text in data documents and the data in datasets. Most commercial applications do not support the creation or maintenance of text-data connections, instead requiring that users maintain these connections in their mind and perform tedious, manual updates to their documents and data. The state-of-the-art research systems that have been created to support the authoring of dynamic and interactive data documents all require the use of programming to specify data bindings~\cite{Conlen2018, Latif2018}, thus posing a higher barrier to entry for novice users. In addition, for each data connection, a user will need to write and update source code to specify and maintain any connections, resulting in tedious workflows, especially for data documents that contain a large amount of data. 

One observation, however, is that the data reported in data documents is naturally embedded with highly descriptive text. These natural embeddings present an interesting opportunity to solve this text-data connection problem in that they may enable systems to infer text-data connections directly from text during one’s writing process. This work thus explores how language-oriented data bindings could be derived from the latent connections that exist between text and data. To systematically explore how language-oriented text-data connections can assist in the authoring of data documents, this research sought to understand the general workflow, pain points, and challenges that exist when authoring data documents by conducting a formative study with eight professionals from different domains who write data documents extensively as part of their daily work. Informed by the findings from this study, we then developed CrossData (\autoref{fig:teaser}), a research prototype that explores the potential of extracting latent language-oriented data bindings that exist within highly descriptive text and reifying them as persistent, interactive, first-class objects~\cite{Instrumentalinteraction, textlet, BeyondSnapping, ood} to assist in the authoring of data documents. 

CrossData utilizes a connection engine that automatically detects, establishes, and maintains text-data connections during the writing process by using state-of-the-art natural language processing (NLP) techniques. While writing text for their documents, CrossData enables users to efficiently retrieve, compute, explore data, and refine tables and charts using interactive interaction techniques that are enabled by the language-oriented data bindings that are identified and created. CrossData leverages these bindings to automatically ensure consistency and congruency between the text, data, tables, and charts. In addition, data documents written with CrossData automatically become interactive documents for readers, enabling them to have a dynamic, explorable reading experience. 
To assess the performance of the connection engine in extracting latent text-data connections, a technical evaluation was conducted. The results showed that the engine correctly constructed 88.8\% of 529 text-data connections identified from 206 sentences, demonstrating its effectiveness.
To assess the utility of language-oriented data bindings, an expert evaluation was conducted and demonstrated that CrossData's interaction techniques can significantly reduce the manual effort required while writing data documents and also enable fluid and enjoyable workflows. Feedback from experts also indicated that language-oriented authoring exposes new possibilities for data exploration and authoring.

This systematic exploration of language-oriented authoring for data documents thus contributes:
\begin{enumerate}[label={\arabic*.}]
    \item An understanding of the challenges that exist when authoring data documents today.
    \item A language-oriented data binding approach that extracts latent text-data connections from written text.
    \item A set of novel interaction techniques that enable users to efficiently author and iterate on data documents.
    \item The CrossData prototype system, i.e., an implementation of language-oriented authoring for data documents, which was evaluated by experts along the dimensions of the usefulness and usability of the interaction techniques that the system supported. 
\end{enumerate}

\section{RELATED WORK}
As this research aims to leverage the connections that exist between highly descriptive text and data to ease the authoring of data documents, prior work on authoring data-driven content, linking text to other visual media, and natural language interfaces for data queries and visualization, are reviewed.

\subsection{Authoring Data-driven Content}
Significant research in HCI and data visualization has explored how to support the authoring of data-driven content, such as charts~\cite{DBLP:conf/chi/ChenTWBQ20, Ren2019},
infographics~\cite{Xia2018, Chen2020}, 
data-driven comics~\cite{Kim2019}, 
videos~\cite{Amini2017},
and articles~\cite{Sultanum2021}. 
Within this research, bindings were created between the visual components and the underlying data so that the data-driven content could be updated whenever the data changed, and vice versa. This thus reduced the repetitive effort necessary to manually update content and enabled rich, dynamic interactive experiences.

There has been a proliferation of research systems that have assisted in the creation of data visualizations that have followed the principles of direct manipulation~\cite{Shneiderman1982, Instrumentalinteraction} as alternatives to the template-based chart editing methods that lack customizability and the programming libraries that require significant expertise and are often cognitively demanding to use. For example, Data Illustrator~\cite{DataIllustrator}, DataInk~\cite{Xia2018}, and Lyra~\cite{Lyra} enabled users to directly create a set of visual encodings, which could be applied to all the data points in a dataset to quickly generate data visualizations. Victor proposed a system that captured parameterized drawing steps, which could later be reused to generate an entire visualization~\cite{Victor2013}. 
Charticulator also allowed authors to interactively specify chart layouts and employed a constraint-based method to realize layouts~\cite{Ren2019}.  

Recent research has extended the concept of data-driven content to other media such as data-driven articles, which consist of text, charts, interactive equations, simulations, and so on. For example, Victor presented Explorable Explanations, a type of data-driven article where the numbers and equations reported in the text were bounded to the underlying data and computation models enabled readers to manipulate the author’s assumptions and see the consequences~\cite{Victor2011}. Dragicevic et al. applied a similar idea to scientific reports, enabling readers to explore the different analytical results of a study~\cite{Dragicevic2019}. 
Computational notebooks (e.g., Jupyter~\cite{Jupyter2021}, R Markdown~\cite{RMarkdown}), an modern embodiments of Knuth’s literate programming notion~\cite{Knuth1984}, also
allowed users to integrate data with text, executable code, and visualizations to reproduce and share explorations.
Creating such data-driven content, however, is tedious and time-consuming because, unlike data visualizations where users can easily configure a small set of visual encodings to create and adjust the entire visualization, each binding in a data-driven article often requires specific configurations with the underlying data. As a result, state-of-the-art systems designed to support authoring data-driven articles use programming languages and require users to manually configure each desired data-driven element. For example, Idyll, a markup language for web-based interactive documents, enabled users to bind data or reader events (e.g., page scrolling) to text, visualizations, and other elements in documents, thereby creating an interactive reading experience~\cite{Conlen2018}.
Computational notebooks require users to write code to manipulate and bind data to other content, while text is mainly used for explanatory descriptions alongside code to facilitate documentation.

Instead of requiring users to manually specify data-driven bindings using programming languages, CrossData infers and recommends connections that implicitly exist between text and data to the user during the writing process. Coupled with a set of novel interaction techniques that enable users to easily select and update text-data connections, CrossData not only significantly reduces the manual effort needed to create data documents, but also simultaneously enables an interactive reading experience for readers without any additional effort.

\subsection{Linking Text to Other Visual Media}
There has been significant research exploring how text can be leveraged and enhanced to facilitate both content consumption and creation processes. 
To facilitate data communication and help users efficiently synthesize information distributed across a data document, prior work has explored connecting text with other data representations (i.e., tables~\cite{Kim2018, Badam2019} and charts~\cite{Sultanum2021, Kong2014, Latif2018}) to enhance reading experiences, using a variety of techniques including direct manipulation, mixed-initiative, crowdsourced, and fully automatic methods. 
For example, Sultanum et al. enabled users to specify desired links between text and charts and leveraged these text-chart links to adapt content to a range of layouts~\cite{Sultanum2021}. 
Latif et al. developed a mixed-initiative interface by leveraging NLP techniques to construct interactive references between text and charts~\cite{Latif2018}. 
Kong et al. developed an interactive document reading application that utilized crowdsourced links between text and charts to enable users to easily navigate from text to referred marks in a chart~\cite{Kong2014}. 
Kim et al. leveraged NLP techniques to connect text with corresponding cells in data tables within PDF documents to enhance reading experiences~\cite{Kim2018}. 
Recent advances in deep neural network have also led to a sequence of automatic methods to facilitate the reading of visualizations with text, such as visualization annotation~\cite{Lai2020}, chart captioning~\cite{autoCaption}, and chart question answering~\cite{Kim2020, DVQA}.

Beyond linking text with different data representations, extensive research in NLP, computer vision, and machine learning has explored the automatic conversion of domain-specific descriptive text into visual content, such as 3D shapes~\cite{Chen2018} and scenes~\cite{Coyne2001, Chang2014}, infographics~\cite{Cui2020}, as well as short video clips~\cite{Marwah2017}, to help content creators. 
For example, the WordsEye system matched word semantics to the functional and spatial properties of 3D models to automatically convert text descriptions into 3D scenes~\cite{Coyne2001}. 
Research in HCI has also leveraged the links between text and visual content to assist in the creation process. For example, Rubin et al.~\cite{Rubin2012} and Troung et al.~\cite{Truong2016} leveraged the linear temporal properties that are common across text, audio, and video to assist in the editing of media clips. 
Perhaps the most closely related work to the present research is Crosspower~\cite{Xia2020}, which leveraged desired correspondences between linguistic structures and graphical structures to enable users to flexibly and quickly create and manipulate graphical elements, as well as their layouts and animations. The present research also seeks to support content creation. However, it focuses on the domain of data documents, which resulted in a different set of interaction techniques to coherently address several challenges in users’ workflows while authoring data documents.

\vspace{-2mm}
\subsection{Natural Language Interfaces for Data Queries and Visualization}
Recent advances in NLP have renewed interest in natural language interfaces (NLIs) for data analysis. Compared to traditional data analysis systems, systems with NLIs enable users to interact with data by using questions and commands expressed via natural language rather than via interface actions or domain-specific languages (e.g., SQL), thereby lowering barriers for non-experts to access data~\cite{Affolter2019}. 
These systems can be roughly divided into two categories based on if they support data queries or if they support the creation of, and interaction with, data visualizations.

Querying data through natural language has been extensively studied in the field of database systems. Many systems from this field adopted a parsing-based strategy~\cite{Ozcan2020, Affolter2019}, with the goal of constructing SQL queries by identifying entities and their relationships in an input query. For example, ATHENA~\cite{ATHENA} parsed and mapped natural language queries to entities in an ontology generated automatically from a database and then translated the input query into SQL. 
Recently, machine learning-based methods have been gaining traction due to the success of deep learning~\cite{Shi2018, Utama2018}. These methods use supervised neural networks to translate a natural language query to SQL. 
Seq2SQL~\cite{Seq2SQL}, for example, used a deep reinforcement learning model to generate SQL based on an input query. To leverage the best of both methods, some systems (e.g., QUEST~\cite{QUEST}) have utilized parsing- and learning-based methods as part of a multi-step pipeline.

NLIs for data visualizations can be seen as an extension of NLIs for databases, which enable users to visualize query results and interact with the generated visualizations. For example, a user can type “show me the medals for hockey and skating by country” to generate a visualization of this specific data. A key challenge when generating visualizations based on natural language is to resolve the ambiguities that exist in the query. 
DataTone~\cite{DataTone}, for example, proposed a mixed-initiative approach that enabled users to resolve ambiguities by interacting with ambiguity widgets. NL4DV~\cite{NL4DV} was a toolkit that took a tabular dataset and a query as input and returned a JSON specification of generated visualizations. Ambiguous results were then highlighted in the specification. In addition to generating visualizations, researchers have also used natural language to interact with visualizations. For example, Eviza~\cite{Eviza} enabled users to continually revise and interact with a visualization by asking questions. 
InChorus~\cite{InChorus} supported multimodal input with both speech and touch to interact with visualizations. 
Recently, Srinivasan et al.~\cite{Srinivasan2021} presented a dataset of visualization-oriented utterances collected from an online study, providing a benchmark of NLIs for visualization.

Overall, these NLI systems treated natural language and text as commands, so there were no persistent connections between the text and the data. While CrossData was built using similar NLP techniques, highly descriptive text was viewed as another representation of the underlying data so it was important to preserve the connections that existed between the text and data. These persistent connections were then leveraged to provide rich interactions that could be used during the writing process.

\section{FORMATIVE STUDY WITH PROFESSIONALS}

To better understand the general workflow, pain points, and best practices while writing data documents, a formative interview study was conducted.

\vspace{-2mm}
\subsection{Participants and Procedure}
Eight professionals from various domains, including business services, e-commerce, accounting, banking, biomedical science, retail, and internet services were interviewed (4 female, age 27 – 30). Each had 3 – 7 years working in their current role and their responsibilities included exploring, analyzing, and reporting data. The interviews were conducted remotely using videotelephony and lasted between 45 to 60 minutes.

During the interviews, the professionals were asked to describe a recent, memorable experience while writing data documents, common pain points, and their solutions. They were also asked to share their documents and tools through screen sharing, if possible. The interview ended with a questionnaire to collect demographic information. Four pilot interviews with another 4 professionals were conducted beforehand to develop the study protocol.

Interviews were audio-recorded, transcribed, and analyzed using a reflexive thematic analysis~\cite{reflexiveTA}. The codes and themes were generated both inductively (i.e., bottom-up) and deductively (i.e., top-down), focusing on the workflow breakdowns, repetitive operations, and workarounds that occurred while writing data documents. 

\vspace{-2mm}
\subsection{Findings and Discussion}

The general process of producing data documents mainly included data exploration and writing. During the exploration stage, participants cleaned, processed, and explored their data with a concrete goal or question assigned to them by their manager. Excel was the most common tool used for this process (7/8). All participants said that when insights and findings were discovered within the data, they would \quot{create or screenshot the table or chart (of the insights), insert it to a Word document, and write a short description for it} (P3). After accumulating enough insights, participants moved to the writing stage. All participants indicated that they frequently revisited the data while writing, as their original insights could be unclear, complicated, incorrect, obsolete, or unappealing to present. Their document would often be iterated on by collaborators, leading to additional data exploration. Thus, their writing processes were highly intertwined with data exploration. Finally, the document would be carefully reviewed together with the data to ensure that there were no inconsistencies between the document and data before delivery.

\vspace{-1mm}
\subsubsection{Tedious and Frequent Data Retrieval (T1)}
\noindent
When writing data documents, participants needed to retrieve data from their data analysis applications (e.g., Excel) to document in the authoring applications they used (e.g., Word). All participants reported that the \quot{frequent application switching and navigation to the data} caused significant friction to the retrieval process. For example, with Excel, participants needed to first identify the correct datasheet, and then scroll within the sheet to locate the data they wanted (P1-6, P8).  Participants (P1-4, P6) often would use the Search function to accelerate their navigation, which required them to memorize specific data properties and navigation pathways when multiple matches were found. Once data was located, participants needed to transfer it to a text editor. While participants often relied on copy-and-paste to avoid errors, they often needed to change the data format (e.g., converting large absolute values to abbreviated forms, P5) or perform simple calculations (e.g., ratio of change, P2), so they had to manually type the data into the document. Each of these steps was tedious but also repeated numerous times during authoring, resulting in time-consuming and error-prone workflows.

\vspace{-1mm}
\subsubsection{Inefficient and Error-prone Maintenance of Data Consistency (T2)}
\noindent
Ensuring consistency between a document and its underlying data was regarded as important, as erroneous data reporting could lead to extra iterations of a document (P2), bad records in one’s career history (P1), or even financial losses for a company (P3). Professionals reported that the inconsistencies were usually caused by data updates. For example, P5, a marketing manager, often started to draft a document before all the data became available so that they could meet deadlines, which led them to  update their analysis and document as soon as new data became available. P3, who worked in a financial services company, frequently updated her documents when there were adjustments in model parameters. Whenever the underlying data was updated, all participants reported that they needed to \quot{read through [their] documents carefully and fix the inconsistent content manually} (P5), which was \quot{inefficient and prone to error} (P1). P1 noted that the IT team in his company developed a plugin that synchronized the data between Excel and Word automatically, but it required the user to manually connect cells to words. P3 mentioned that a professional review team in her company would proofread her documents to highlight any inconsistencies. Nevertheless, these methods were noted as being cumbersome, expensive, and time-consuming.

\begin{figure*}[ht]
  \centering
  \includegraphics[width=0.8\linewidth]{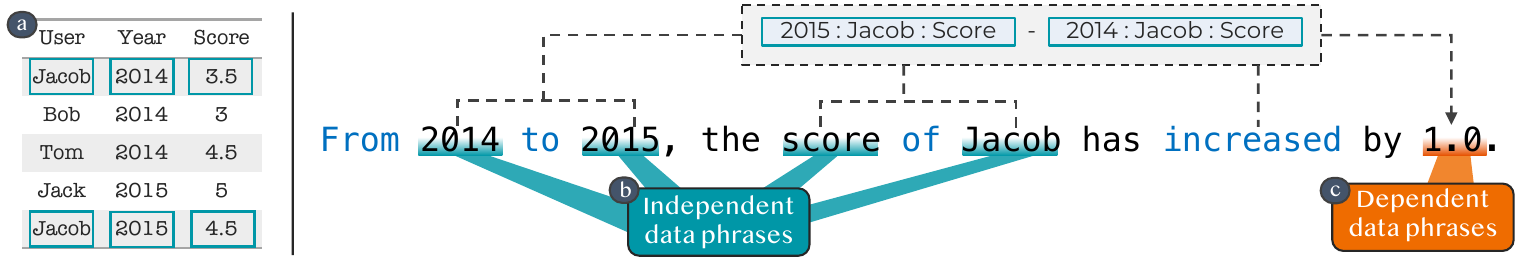}
  \caption{The connections between text and data. a) The dataset to report. b) Data phrases directly reporting the underlying data.
  c) A data phrase connecting with the data under the constraints of other phrases. The Blue text represents the keywords used to compute dependent phrases. }
  \Description{The connections between text and data. a) The dataset to report. b) Data phrases directly reporting the underlying data.
  c) A data phrase connecting with the data under the constraints of other phrases. The Blue text represents the keywords used to compute dependent phrases. }
  \label{fig:dp_example}
\end{figure*}

\vspace{-1mm}
\subsubsection{Significant Overhead for Iteration (T3)}
\noindent
Participants reported that exploring different ways to present data was a common but time-consuming task (7/8). They needed to perform additional data exploration during the writing stage, because \quot{only when I write down the data in the document, I know what's the best way to present it} (P2). As an operating officer in an IT company, P2 reported that she needed to frequently switch the presentation of user growth data on a yearly, quarterly, and monthly basis. 

Exploring alternative data presentations, however, was reported as being time-consuming, because participants often needed to repeat their analysis steps, create new tables and charts, and update the relevant text with new data. P6 mentioned she always used tables or charts to show evidence for the insights reported in the text, i.e., \quot{if I want to report a new metric, I will add one more column to the table} (P6). P8 noted that to \quot{add one more sentence} to introduce \quot{the ratio of a group of users to all users}, he needed to go back to Excel, perform numerous operations to re-create tables and charts, and then insert them into the document. 

Participants reported that during the writing stage, they frequently iterated on the presentation of data. However, even the smallest changes caused significant ripple effects to the data reported in the text, as well as the corresponding tables and charts. Due to such significant overhead, participants and their collaborators had to iterate on the document offline when iterations were suggested in real-time, requiring additional meetings and discussions, thus hindering their collaborative process.

\vspace{-2mm}
\subsection{Summary}
The formative study found that professionals encountered several issues while writing data documents with mainstream tools and they addressed these issues manually. They struggled while inputting the data into their documents, maintaining the consistency between their documents and data, and handling the numerous interconnected components during iterations. The findings indicate that the key reason for their tedious and ineffective workflows was the lack of connections that existed between the text in data documents and the data in datasets, which needed to be created and maintained with minimal effort from users.

\section{CrossData}
When using text to describe data from a dataset in a document, a user establishes an abstract connection between the text and the data elements in their mind. A key insight from the formative study was that current tools require the user to mentally maintain these connections, leading to tedious, repetitive, and error-prone operations. We propose reifying these connections as persistent, first-class objects~\cite{Instrumentalinteraction, textlet, BeyondSnapping} and leveraging them to address the issues that occur during the writing process. To this end, two steps were undertaken: 1) we developed a connection engine to automatically establish and maintain these connections during writing processes and 2) we designed a set of interactions based on these connections to tackle the issues identified in the formative study. The present work focuses on tabular data, which is one of the most common data formats.

\section{The CONNECTION ENGINE for text-data connections}
Given the text in a data document and an underlying dataset, our goal was to infer, establish, and maintain text-data connections. 

\subsection{Connections Between Text and Data}
When describing data using text, the phrases in text can connect with the underlying data in two ways:

\begin{enumerate}[label={\arabic*.}, leftmargin=*]
    \item \textbf{Independent data phrases}, directly report items (rows), attributes (columns), and values (cells) in the dataset. For example, in~\autoref{fig:dp_example}, \emph{2014}, \emph{2015}, \emph{score}, and \emph{Jacob} (\autoref{fig:dp_example}b) are connected to the cells in the table (\autoref{fig:dp_example}a). Independent data phrases can be used as arguments to compute dependent data phrases.
    \item \textbf{Dependent data phrases}, present the output of data operations that take other data phrases as arguments. A dependent data phrase can report data in the dataset or derived values that do not exist in the dataset. For instance, the last term \emph{1.0} (\autoref{fig:dp_example}c) is calculated based on the other phrases and connects to the data dependently. The data operations to compute a dependent data phrase are described by keywords such as \emph{from}, \emph{to}, and \emph{increased}. 
\end{enumerate}

\subsection{Establishing Text-Data Connections}
\begin{figure*}[ht]
\vspace{-2mm}
  \centering
  \includegraphics[width=0.85\linewidth]{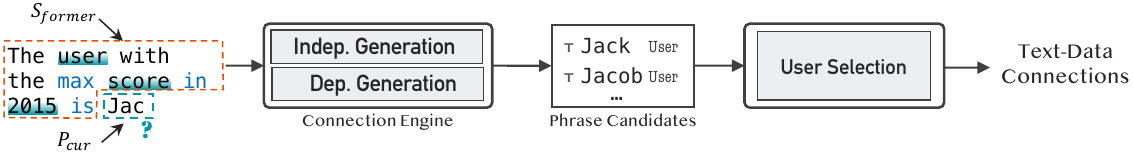}
  \vspace{-2mm}
  \caption{The pipeline to establish text-data connections. The Connection Engine takes a sentence as input and outputs a list of data phrase candidates. The user can select from the candidates to establish text-data connections.}
  \Description{The pipeline to establish text-data connections. The Connection Engine takes a sentence as input and outputs a list of data phrase candidates. The user can select from the candidates to establish text-data connections.}
  \label{fig:pipeline}
  \vspace{-2mm}
\end{figure*}

\begin{figure*}[hb]
  \centering
  \includegraphics[width=\linewidth]{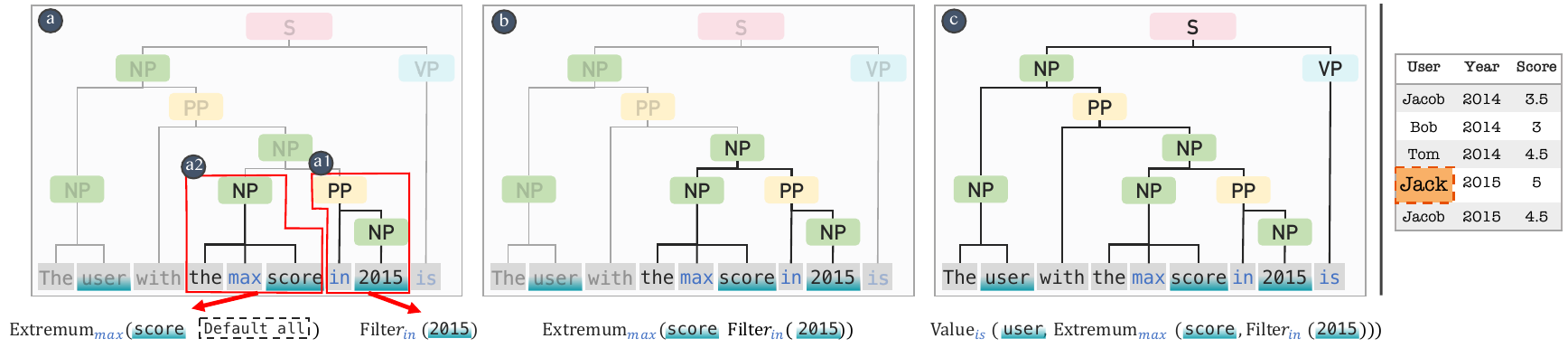}
  \vspace{-4mm}
  \caption{An example detailing how the Connection Engine infers the data operations and suggests dependent data phrases. The engine first parses the sentence into a constituency tree, each of whose nodes represents text phrases (e.g., noun/verb/proposition phrase) in the sentence. Then, the engine infers and assembles data operations in a bottom-up order (a - c). The output of the operation in the root node is returned as suggested dependent data phrases.}
  \Description{An example detailing how the Connection Engine infers the data operations and suggests dependent data phrases. The engine first parses the sentence into a constituency tree, each of whose nodes represents text phrases (e.g., noun/verb/proposition phrase) in the sentence. Then, the engine infers and assembles data operations in a bottom-up order (a - c). The output of the operation in the root node is returned as suggested dependent data phrases.}
  \label{fig:cExample}
\end{figure*}

\vspace{-1mm}
\noindent
The Connection Engine helps users establish and maintain connections during the writing process (\autoref{fig:pipeline}). 
Suppose that after writing the first half of a sentence ($S_{former}$), a user is typing a new phrase ($P_{cur}$). The Connection Engine generates all potential connections for $P_{cur}$, which are presented as a list of data phrases to the user. 
Once a data phrase is chosen by the user, the Connection Engine inserts the phrase into the document with the text-data connection and all relevant meta information is maintained.

\vspace{-1mm}
\subsubsection{Establishing Connections for Independent Data Phrases}

\noindent
The Connection Engine generates the potential independent phrases for $P_{cur}$ by performing string matching of $P_{cur}$ with all strings in the dataset and synonym matching with all attribute names in the dataset. The synonym matching is achieved by calculating the similarity of the word embeddings provided by Spacy~\cite{spacy}, an industrial-strength NLP toolkit. 
All matches will then be returned as suggestions, ordered by their matching scores. Selecting a suggestion will insert an independent phrase and create a connection between the independent phrase and the underlying dataset.

\vspace{-1mm}
\subsubsection{Establishing Connections for Dependent Data Phrases}

\noindent
Since dependent data phrases are the result of data operations that take other phrases as arguments, the Connection Engine takes three steps to identify, assemble, and execute the data operations, and then returns the results of the data operations as suggestions to the user. Selecting a suggestion will insert a dependent data phrase and establish a connection with the underlying data operation:

\begin{enumerate}[label={\arabic*.}, leftmargin=*]
    \item \textbf{Identifying data operations}: To detect data operations, the Connection Engine matches words and phrases with keywords in a predefined operation dictionary. The dictionary is derived from Amar et al.’s work, which summarized 10 low-level analytical operations for data analysis, such as retrieve value, filter, and compute derived value~\cite{Amar2005}. This summarization has been widely used in NLI systems to extract desired data operations from users’ input queries~\cite{DataTone, NL4DV}.
    An operation takes a few arguments as input and outputs either an item (row), an attribute (column), a value (cell), or a derived value of the underlying dataset. The detailed definition of operations implemented in the current system is provided in the supplemental materials.
    
    \item \textbf{Assembling data operations with arguments}: As an operation needs arguments to compute output, the arguments of an operation can either be independent data phrases or the output of other operations. To infer the arguments for each operation, we parse the input text as a constituency tree using the Berkeley Neural Parser~\cite{Kitaev2019} through its integration with Spacy. Within a constituency tree, each node represents a text phrase in the sentence (e.g., noun/verb/proposition phrases), with smaller phrases being deeper in the tree, i.e., the leaf nodes are words. Therefore, the Connection Engine uses a bottom-up order to recursively examine whether the independent data phrases and operations in a node can be assembled as a complete data operation, as well as whether data operations should be assembled as compounded data operations. The Connection Engine employs a rule-based method to achieve the examination, as explored in previous NLI research~\cite{NL4DV, DataTone, Eviza}.
    Specifically, the Connection Engine matches the set of phrases and their grammatic relationships (also provided by Spacy) of a node with pre-constructed rules, each of which describes the necessary arguments for a data operation and the required data types (i.e., item, attribute, or value) for the arguments. The pseudocode for the assembling process is provided in the supplemental materials.
    
    \item \textbf{Executing data operations}: Finally, the Connection Engine executes the data operation in the root node of the sentence to obtain the result. Since a keyword may match different operations, the Connection Engine employs a greedy strategy to enumerate all possible matched operations for a keyword, assemble them into complete operations, and return all the results as dependent phrase candidates for the user.
\end{enumerate}

\begin{figure*}[ht]
  \centering
  \includegraphics[width=0.85\linewidth]{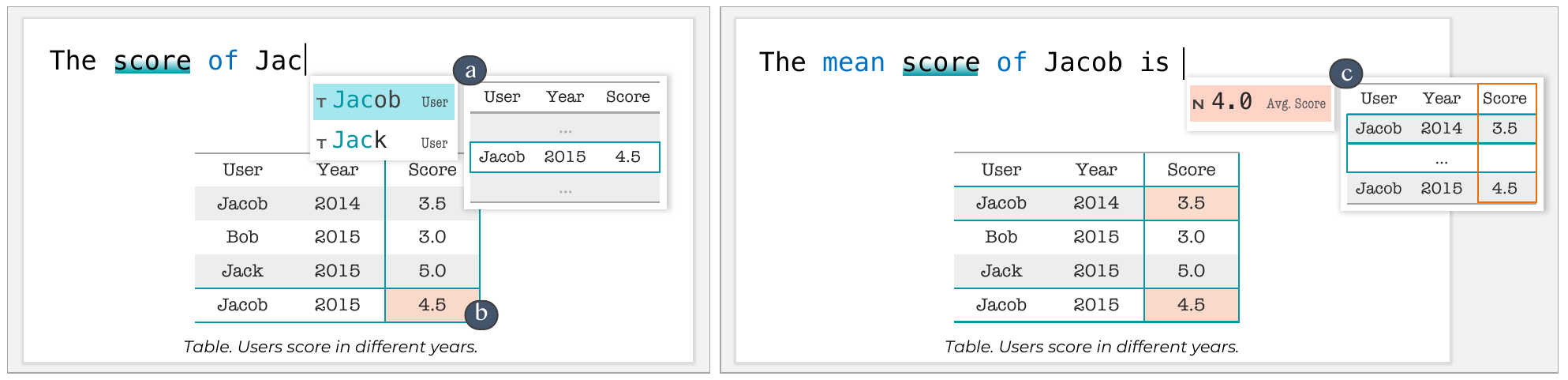}
  \vspace{-2mm}
  \caption{Retrieving Data and Computing Values. a) A list of independent data phrases (highlighted by the cyan background) are retrieved and suggested for the user. b) The data mentioned in the sentence is highlighted. c) The mean score is computed and suggested as a dependent data phrase (highlighted by the orange background) for the user. Detail information about each suggestion is provided to assist in resolving ambiguities.}
  \Description{Retrieving Data and Computing Values. a) A list of independent data phrases (highlighted by the cyan background) are retrieved and suggested for the user. b) The data mentioned in the sentence is highlighted. c) The mean score is computed and suggested as a dependent data phrase (highlighted by the orange background) for the user. Detail information about each suggestion is provided to assist in resolving ambiguities.}
  \label{fig:input}
\end{figure*}

Take the sentence \quot{The user with the max score in 2015 is} as an example. 
The Connection Engine starts the inferring process from the leaf node \quot{2015}, which reports a value in the data. 
Since \quot{2015} is an independent phrase and the only one at the lowest level, no data operations can be inferred. 
The Connection Engine then recursively processes the parent nodes of 2015 to a proposition phrase (PP) node and infers a filter operation for the keyword \quot{in} with \quot{2015} as the argument (\autoref{fig:cExample}a1). 
Similarly, the Connection Engine infers a find extremum operation for the keyword \quot{max} on the \quot{Score} column from the phrase \quot{the max score} (\autoref{fig:cExample}a2). 
According to our predefined rules, the operation finds the extremum in all rows by default. When process to its parent node (\autoref{fig:cExample}b), the engine fills the default argument (i.e., all rows) with the output of the filter operation inferred in \autoref{fig:cExample}a1 since its output is a list of rows.
The engine recursively repeats this process and finally infers a retrieve value operation in the root node from the keyword \quot{is}, 
whose arguments are the phrase \quot{user} and output of the find extremum operation (\autoref{fig:cExample}c). 
As such, the dependent data phrase is computed from a compounded operation of the filter, find extremum, and retrieve value operations. 
The output of this compound operation, \quot{Jack}, will then be recommended to the user.
Once the user selects \quot{Jack} from the suggestions, a dependent phrase will be inserted, and a text-data connection will be established. 
The detailed rules for each operation and the pseudocode of the algorithm are provided in the supplemental materials.

\section{LEVERAGING CONNECTIONS for DATA DOCUMENT AUTHORING}
CrossData leverages the text-data connections found by the Connection Engine to provide novel interactions that address the issues identified in the formative study, thus enabling users to efficiently retrieve, compute, explore data, and adjust tables and charts during the writing of data documents, while automatically maintaining data consistency between the text, data, tables, and charts.

\begin{figure*}[hb]
  \centering
  \includegraphics[width=0.9\linewidth]{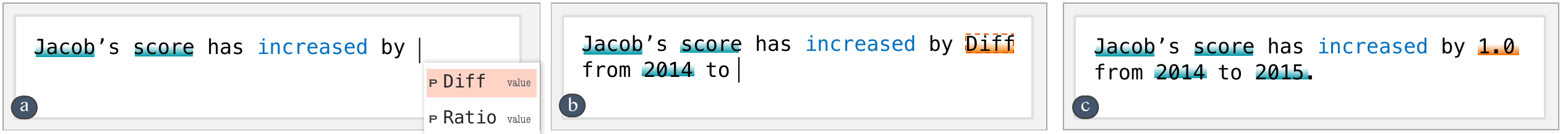}
  \caption{Using placeholders. a) There is not enough information provided in the sentence to calculate the difference between Jacob’s scores in different years. b) CrossData allows the user to use a Diff placeholder to indicate the computation. c) CrossData updates the placeholder as more information is provided.}
  \Description{Using placeholders. a) There is not enough information provided in the sentence to calculate the difference between Jacob’s scores in different years. b) CrossData allows the user to use a Diff placeholder to indicate the computation. c) CrossData updates the placeholder as more information is provided.}
  \label{fig:placeholder}
\end{figure*}

\subsection{Connections for Inputting Data}

The formative study found that data retrieval is tedious but repeated numerous times when authoring data documents (T1). Professionals manually retrieved data from data processing tools (e.g., Excel), leading to issues while application switching, navigating data, and transferring data into word processing tools (e.g., Word). To address these issues, several interactions that enable users to leverage the output of the Connection Engine were thus designed.

\subsubsection{Retrieving Data}

\noindent
As a user types in the text editor, CrossData automatically runs the Connection Engine to detect the connections. The underlying data elements that the text potentially connects to are returned as suggestions for the user in a list (\autoref{fig:input}a). 
Additional information (e.g., the data types, the context in the spreadsheet, etc.) about each suggestion is provided for each list item to help the user select the correct data and resolve ambiguities. 
If the underlying data table is also visible on the user interface, CrossData automatically highlights the corresponding row, column, or cell based on the data phrases the user is typing (\autoref{fig:input}b).
Such reference highlighting can help users efficiently locate the elements in tables.
The user can select a suggestion from the list to insert it into the text editor or simply enter the text following the suggestion.
CrossData will automatically maintain the connection between the text and data for later reuse.

\subsubsection{Computing Values}
\noindent
Sometimes the user needs to compute and input values that do not exist in their dataset. 
CrossData detects these dependent connections and calculates their derived value using the Connection Engine. The derived value and the detailed information about the calculation are displayed as suggestions for the user (\autoref{fig:input}c).
The user can select and insert the derived data while preserving the connection.

\subsubsection{Using Placeholders}

\noindent
An issue when retrieving or computing data in a written sentence, which differs from command-like sentences in other NLIs systems, is that the data that one may want to retrieve or compute could be input before its dependency is retrieved or computed. CrossData thus provides a set of placeholders, such as Diff, Ratio, and Count, that the user can employ to indicate expected data types. 
For example, in \autoref{fig:placeholder}a, if the user wants to report the increase in Jacob’s score while the year range is unknown, the user can press the Tab key to open the suggestion list to select and insert a placeholder (\autoref{fig:placeholder}b). 
Then, whenever new data phrases in the sentence are inserted or detected, the Connection Engine will attempt to evaluate and update the placeholders (\autoref{fig:placeholder}c). 
All placeholders are thus dependent data phrases.

\subsubsection{Fixing Misdetections}

\begin{figure}[h]
  \centering
  \includegraphics[width=0.9\linewidth]{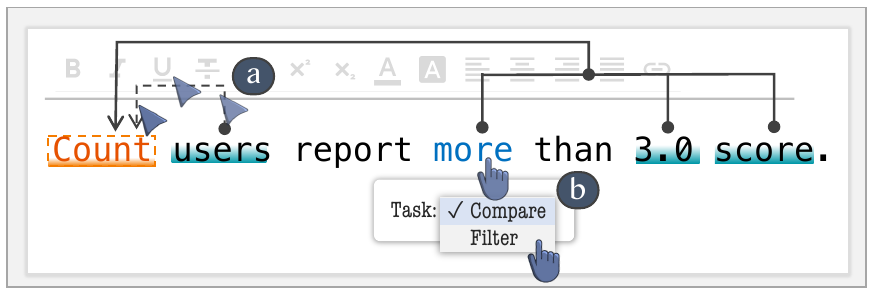}
  \vspace{-2mm}
  \caption{Fixing misdetections by a) hovering over the \quot{Count} placeholder to visualize its dependencies and linking \quot{users} to \quot{Count} to fix the missing dependency or by b) hovering over the operation keyword \quot{more} to display the task inferred from it. In this example, \quot{more} should be interpreted as a filter instead of a comparison task.}
  \Description{Fixing misdetections by a) hovering over the Count placeholder to visualize its dependencies and linking users to Count to fix the missing dependency or by b) hovering over the operation keyword more to display the task inferred from it. In this example, more should be interpreted as a filter instead of a comparison task.}
  \label{fig:fixmis}
\end{figure}

\noindent
It is not uncommon for CrossData to retrieve or calculate incorrect data for dependent data phrases. The incorrectness can be caused by mis-detected dependencies (i.e., wrong input) or operation keywords (i.e., wrong tasks). 
CrossData allows the user to interactively correct these misdetections by hovering over a dependent data phrase to visualize and modify its dependencies (\autoref{fig:fixmis}a) or hovering over operation keywords to refine their tasks (\autoref{fig:fixmis}b).

\subsection{Connections to Maintain Consistency}

\begin{figure}[h]
  \centering
  \includegraphics[width=\linewidth]{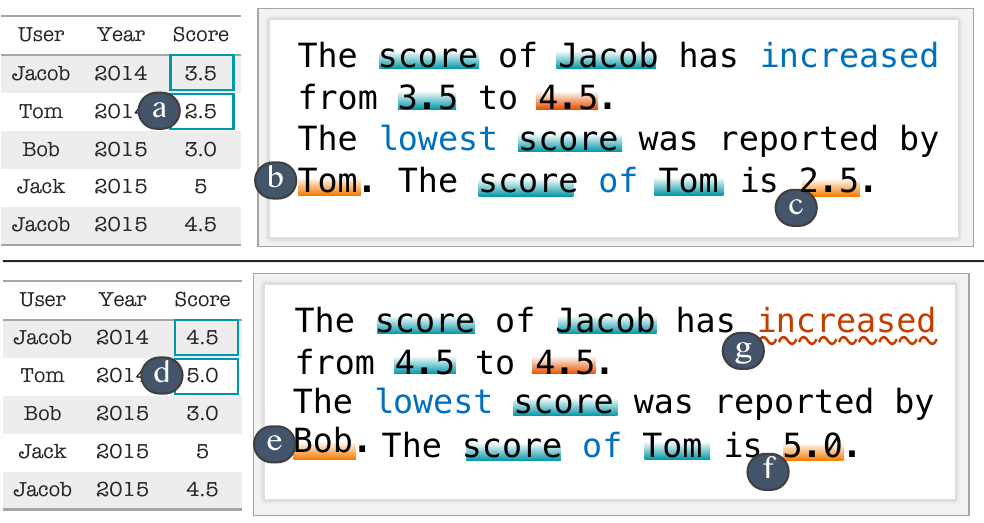}
   \vspace{-2mm}
  \caption{Maintaining consistency automatically. After changing Tom’s score (a) from 2.5 to 5.0 (d), CrossData updates all related sentences, such as the user with the lowest score (b) and Tom’s score (c). Problematic operation keywords caused by the updated data will also be highlighted, e.g., after changing Jacob’s score (a) from 3.5 to 4.5 (d), the \quot{increase} description (g) is incorrect.}
  \Description{Maintaining consistency automatically. After changing Tom’s score (a) from 2.5 to 5.0 (d), CrossData updates all related sentences, such as the user with the lowest score (b) and Tom’s score (c). Problematic operation keywords caused by the updated data will also be highlighted, e.g., after changing Jacob’s score (a) from 3.5 to 4.5 (d), the increase description (g) is incorrect.}
  \label{fig:consistency}
\end{figure}

\noindent
The formative interviews demonstrated that most of the professionals manually maintained consistency between their text and data and considered this process to be time-consuming and error-prone (T2). With the help of preserved connections, CrossData can update data phrases and highlight problematic operation keywords to help users maintain consistency. 

\subsubsection{Data-driven Updates}
\noindent
Whenever a data element within the underlying dataset is updated, CrossData will automatically update all independent and dependent phrases that connect to the data element. 
For example, if the user changes the score of Tom from 2.5 (\autoref{fig:consistency}a) to 5.0 (\autoref{fig:consistency}d) in the table, 
CrossData will update Tom’s score to 5.0 (\autoref{fig:consistency}f) in the last sentence; meanwhile, Tom (\autoref{fig:consistency}b) will be updated to Bob (\autoref{fig:consistency}e) accordingly.

\subsubsection{Operation Keywords Checker}
\noindent
Inconsistencies can also exist between the operation keywords and the data.
For example, when changing the score of the first row from 3.5 (\autoref{fig:consistency}a) to 4.5 (\autoref{fig:consistency}d), 
the operation keyword \quot{increase} is inconsistent with the data.
However, different from data phrases, updating operations can be challenging because operation phrases are usually text descriptions. In such cases, CrossData will highlight the problematic operation keyword with red wavy underline (\autoref{fig:consistency}g).

\subsection{Connections for Interactive and Flexible Iteration and Exploration}
When iterating on a data document, users frequently change various elements in their document (T3). While the interaction techniques introduced above can alleviate the overhead of retrieving values and maintaining consistency during iteration, a pressing and unaddressed challenge is the cascading effects that occur when changes are made to text, tables, and charts. 

CrossData addresses this challenge by reifying text-data connections as interactive objects, which enable users to manipulate them to iterate on data documents and explore new insights directly in a document. Because the data phrases, tables, and charts are all connected with the underlying data, the necessary changes can be automatically performed without additional user effort.

\subsubsection{Interacting with Data-Driven Text}
\begin{figure}[b]
  \centering
  \includegraphics[width=0.85\linewidth]{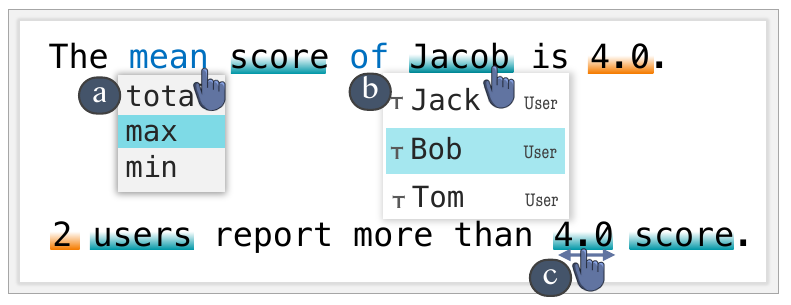}
   \vspace{-2mm}
  \caption{Interactive text. CrossData enables users to interactively iterate operation keywords (a) and independent (b, c) phrases. The interactions will trigger the related dependent data phrases to be updated.}
  \Description{Interactive text. CrossData enables users to interactively iterate operation keywords (a) and independent (b, c) phrases. The interactions will trigger the related dependent data phrases to be updated.}
  \label{fig:interactive}
\end{figure}

\noindent
Text phrases that are connected with underlying data can be interactively manipulated. As independent phrases represent an item (row), attribute (column), or value (cell) within the spreadsheet, CrossData allows the user to interactively change an independent phrase to other items, attributes, or values (\autoref{fig:interactive}b, c).
The interactions provided by an independent phrase depend on its data type, e.g., quantitative, nominal, or ordinal. To avoid meaningless changes, CrossData only allows users to change item phrases to other items, attribute phrases to other attributes that have the same data type, and value phrases to other values in the same column.

Users often need to iterate on the metrics they use to report on their data, such as changing the average value to the median value or from a daily basis to a weekly basis. 
CrossData enables users to interactively alter operation keywords to achieve such goals. For example, in~\autoref{fig:interactive}a, the user can click and change the mean to other computations such as total, maximum, or median. The available operation keyword alternatives are predefined within a curated dictionary.

Changes to interactive text phrases are automatically propagated to other phrases according to the inferred data operation. For example, in~\autoref{fig:interactive}b, if the user interactively changes \emph{Jacob} to \emph{Bob}, CrossData will update the value \emph{4.0} to Bob’s mean score.

\subsubsection{Automatic Adjustments of Tables and Charts}

\begin{figure*}[hb]
  \centering
  \includegraphics[width=\linewidth]{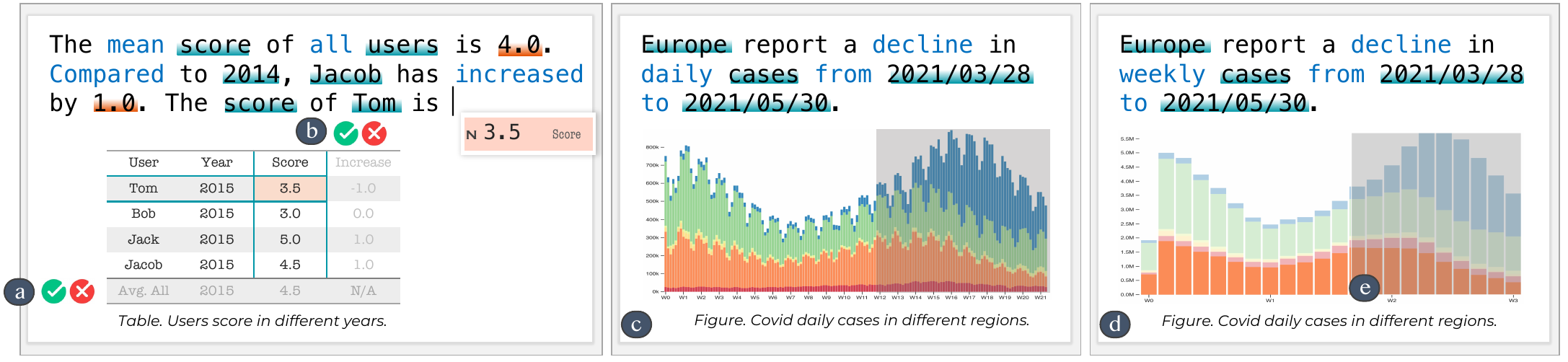}
   \vspace{-2mm}
  \caption{Adjustments of tables and charts based on the text. Based on the user’s writing, CrossData adds a new row and new column to the table (a) and switches the data sources from daily (b) to weekly (d). Users can also directly manipulate the chart to update the text, e.g., by dragging the chart annotation (e).}
  \Description{Adjustments of tables and charts based on the text. Based on the user’s writing, CrossData adds a new row and new column to the table (a) and switches the data sources from daily (b) to weekly (d). Users can also directly manipulate the chart to update the text, e.g., by dragging the chart annotation (e).}
  \label{fig:adjustment}
  \vspace{-2mm}
\end{figure*}

\noindent
Because the text, tables, and charts embedded in a document are all connected to their underlying data, CrossData can automatically update tables and charts with the text to ensure the textual descriptions and data visualizations are consistent. CrossData supports three types of language-oriented manipulations of embedded data tables, based on the detected data operations in the text. First, when a dependent phrase is the output of a sort or find extremum task, CrossData will sort the table based on the column involved in the task. Second, if the user computes a dependent phrase by aggregating multiple rows (e.g., summation), CrossData automatically adds a new row that shows the aggregation results to the table (\autoref{fig:adjustment}a). Third, if the dependent phrase computes a new attribute for an item (e.g., the increase from last year), CrossData will attempt to calculate this attribute for all rows and add a new column to the table (\autoref{fig:adjustment}b). Changes in the tables are suggested to the user, which they can accept or reject.

Similarly, embedded charts are also synchronized with textual descriptions. CrossData automatically updates the charts if different data properties are reported in the text. For example, when the user switches the reporting of new infection cases from daily (\autoref{fig:adjustment}c) to weekly (\autoref{fig:adjustment}d), CrossData will automatically switch the underlying data source of the chart to synchronize with the change. 
CrossData will also automatically annotate the time period of the charts based on the dates reported in the text (\autoref{fig:adjustment}e).
Since both the text and chart are connected to the underlying data, the user can directly manipulate the chart to adjust the text (e.g., dragging the chart overlay in \autoref{fig:adjustment}e), or vice versa, which can facilitate better authoring and reading experiences.

The supported editing operations are limited in the current implementation because the scope of this work is to demonstrate promising novel interactions and workflows enabled by text-data connections. Nevertheless, it is possible to extend CrossData to support more visualization editing operations and this is left for future exploration.

\section{TECHNICAL EVALUATION OF THE CONNECTION ENGINE}
The effectiveness of CrossData depends on whether the Connection Engine can suggest the correct data phrases to the user. Therefore, we conducted a technical evaluation to assess the accuracy and robustness of the Connection Engine. We report on the experiment settings, results, and investigation of the failure cases, as well as potential improvements to the Connection Engine.

\subsection{Experiment Settings}

\subsubsection{Methodology}
\noindent
The goal of the evaluation is to assess whether the Connection Engine can suggest the correct data phrases based on the text in the writing process. Because independent data phrases are suggested based on string matching, which is usually highly accurate, we focused on evaluating the generation of dependent data phrases. Specifically, we gathered a corpus of sentences together with their corresponding datasets. For each sentence, we manually labelled all independent data phrases with the connections to the datasets as part of the input and all dependent data phrases as ground truth. We then input each sentence word by word into the Connection Engine to simulate a realistic writing experience and compared the suggested dependent phrases against the ground truth. The experiment was run on a Macbook Pro with a i7 2.2GHz Intel CPU.

\subsubsection{Dataset}
\noindent
We collected sentences from 10 data documents from reputable public sources that cover multiple domains,
such as World Health Organization~\cite{WHO}, Bureau of Labor Statistics~\cite{BureauLabor}, 
Pew Research Center~\cite{Gramlich2021}, 
National Center for Education Statistics~\cite{EducationCenter}, 
National Institutes of Health~\cite{NIH2021},
California Department of Public Health~\cite{California},
and a private company~\cite{bilibili},
as well as their corresponding datasets. We sampled the sentences by: 1) manually filtering all sentences that reported data in the documents and then 2) randomly sampling no more than 30 sentences from each document. 
For each sentence, we manually labeled the independent and dependent phrases.
In total, the corpus contained 206 sentences (5398 words), with 807 independent phrases and 529 dependent phrases.

\vspace{-2mm}
\subsubsection{Metrics}
\noindent
We measured the ratio of correct dependent data phrases recommended by the Connection Engine to the total number of dependent data phrases. When the engine returned multiple candidates for a dependent phrase, we counted it as correct if the top 5 candidates contained the correct one. We also measured the time to compute the candidates.

\subsection{Results}
The accuracy of the dependent phrases was 88.8\% (i.e., 470 corrects), which demonstrates the robustness and accuracy of the Connection Engine. Among these correct cases, the majority were computed by the compounded operation of filtering and retrieving values (i.e., 262 cases, 55.7\%), the finding extreme operations (i.e., 62 cases, 13.2\%), the compounded operation of finding extreme operations and retrieving values (i.e., 61 cases, 13.0\%), and the compounded operation of finding extreme operations and comparing values (i.e., 48 cases, 10.2\%). 
This echoes the findings from Section 3.2, reflecting that the data retrieval operation was prevalent in real-world data documents. The average time to generate candidates was 0.3 seconds, which was sufficient for interactive use cases and could be further optimized with 
better implementations.

\subsection{Failure Cases Analysis}
\label{ssec:failure}
We further investigated the failure cases and identified three major reasons for these failures. Note that a failure may be caused by multiple factors.

\subsubsection{E1: Lack of Context (i.e., 50.8\% of cases)}
\noindent
Among the failure cases, the majority of cases (i.e., 31) failed because certain expressions (e.g., it, these, previous years) referred to other data phrases. For example, with the sentence \quot{These three countries comprised 89\% of all cases reported in the region}, to compute the \quot{89\%}, the Connection Engine needed to know which countries \quot{These three countries} referred to. 
In this example, the three countries were mentioned in previous sentences as independent phrases. This problem, however, can be addressed by employing co-reference resolution, i..e, finding expressions that refer to the same entity within or between sentences, which has been advanced in recent years. 
The Connection Engine can integrate co-reference resolutions models~\cite{Lee2017} to connect data phrases in previous sentences to the present one, thereby maintaining the context to infer text-data connections.

\subsubsection{E2: Expect Textual instead of Numerical Outputs (i.e., 27.9\% of cases)}
\noindent
Seventeen cases failed because the expected output was a text description rather than a number. For example, in \quot{Two in five e-cigarette users reported usually paying for their own e-cigarettes}, the expected output was \quot{Two in five} while the engine returned \quot{43\%}.
To address this issue, the Connection Engine could generate more candidates with different formats, or adopt more advance generative language models, such as GPT-3~\cite{Brown2020}. 
Note that while the data formats of the suggested phrases do not match the ground truth, the underlying data operations inferred by the Connection Engine were correct. 
This means that the Connection Engine could accurately infer 91.9\% of all data operations.

\subsubsection{E3: Uncovered Operations (i.e., 21.3\% of cases)}
\noindent
Thirteen cases failed because the required data operations in the sentences were not covered by the 10 low-level data operations summarized by Amar et al. ~\cite{Amar2005}. 
In the example \quot{Cases have decreased steeply for the past four weeks}, computing the \quot{four weeks} is a high-level analytical task (i.e., given a column and a text description of the trend, report the range of rows that fulfill the trend), that is not supported in our prototype. 
Considering the rule-based nature of the Connection Engine, these cases could be addressed by extending the predefined operation dictionary and corresponding rules.

\subsection{Summary}
In summary, the analysis showed that the Connection Engine was robust enough to achieve a high accuracy when generating dependent phrases about a set of real-world sentences collected from multiple domains. The in-depth analysis indicated that most of the failure cases could be fixed by extensions to the engine.
\section{EXPERT EVALUATION}

We developed CrossData as a technology probe to explore the notion of language-oriented data bindings and recognized that it may, at first, create usability problems for users who are familiar with existing tools. To gain feedback about the effectiveness of our approach without being bogged down by the initial challenges some users may encounter with usability, we conducted an expert evaluation study. We focused on collecting experts’ feedback about the usefulness of each interaction technique and how language-oriented authoring could facilitate the overall workflow of authoring data documents.

\subsection{Participants and Apparatus}
Eight participants were recruited to participate in the study (E1 – E8, 5 female, age 28 – 31), i.e., 1 auditor (accounting), 1 operation officer (internet services), 1 investment banking associate (financial services), 1 due diligence consultant (business services), 2 marketing managers (internet services and retail) and 2 researchers (data science and public healthy). 
E1-E5 participated in the formative study. All participants had more than 5 years of experience analyzing data and writing data documents as part of their daily work. The most used data processing and writing tools included Microsoft Excel, Sheets, Word, Google Docs, and Tableau. 
The study was conducted remotely with CrossData implemented as a responsive Web application that participants could directly access from their personal computers. 
Video conferencing was used to communicate with participants, share screens, and record the study. Participants received \$60 (USD) for the approximately 90-minute session.

\subsection{Procedure}
Each session included the following phases:

\subsubsection{Introduction and Training (30 mins)}
\noindent
The experimenter first introduced the study protocol, research motivation, and concepts of CrossData. 
Then, the experimenter walked the participants through the system with an example that contained two datasets that were presented as a table and a bar chart, and five insights to report. Participants were encouraged to ask questions anytime during the process. Participants were then asked to replicate the example to become familiar with the system.

\subsubsection{Reproduction Task (15 mins)}
\noindent
Participants were asked to reproduce a given data document, which presented a USA COVID-19 dataset with a multiple line chart and six sentences, each of which reported an insight. The original datasets, a multiple line chart, and a choropleth map were provided as the context for the insights. 

\subsubsection{Creation Task (20 mins)}
\noindent
Participants were asked to write a short document to report on three datasets about Global COVID-19 cases. Each dataset included one data representation (i.e., a chart or a table) and three insights. The short document needed to contain at least one insight from each dataset, and one data representation. To simulate realistic iterative processes, after the participants finished the document, the experimenter asked them to iterate on the document by 1) reporting two more insights, 2) inserting one more chart or table, and 3) changing the data phrases or operators in the documents. The changes to the data phrases or operators were selected to ensure that the participants experienced all of the proposed interaction techniques.

\begin{figure*}[hb]
  \centering
  \includegraphics[width=\linewidth]{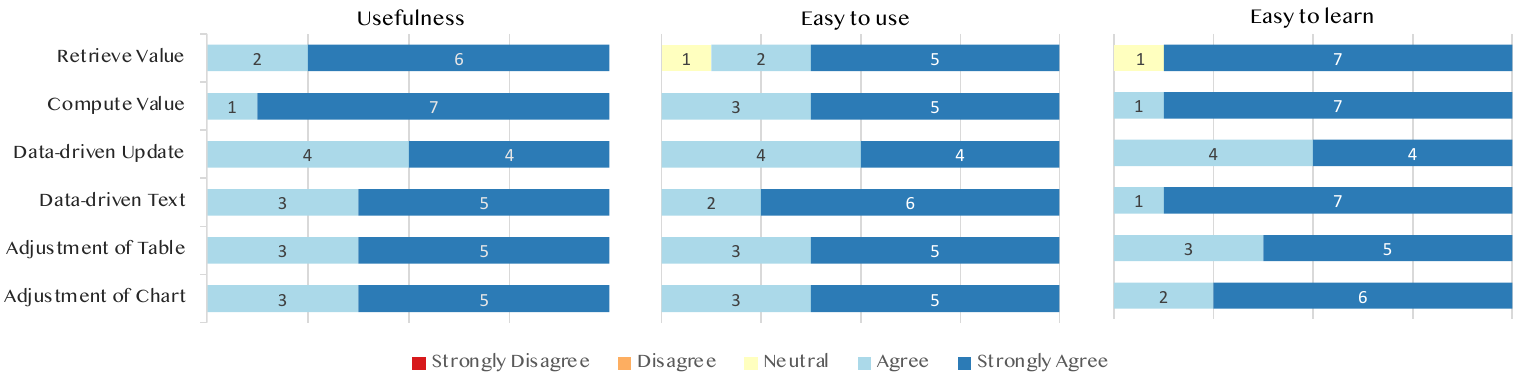}
  \caption{Likert-scale responses to ``This technique was useful in my writing process.'', ``This technique was easy to learn.'', and ``This technique was easy to use.''}
  \Description{Likert-scale responses to ``This technique was useful in my writing process.'', ``This technique was easy to learn.'', and ``This technique was easy to use.''}
  \label{fig:study}
\end{figure*}

\subsubsection{Semi-structured Interview and Questionnaire (25 mins)}
\noindent
After the creation task, participants completed a questionnaire that probed the usefulness and usability of the techniques using a 5-point Likert scale (i.e., 1 – Strongly Disagree, 5 – Strongly Agree). Then, the experimenter conducted a semi-structured interview to further collect feedback about the utility of each interaction technique, CrossData’s effectiveness in supporting realistic workflows, limitations of the proposed techniques, and potential improvements.

\subsection{Results}

All participants successfully finished the reproduction and creation tasks. On average, each participant wrote 12.6 sentences and 123.3 words, which contained 22.1 independent and 13.6 dependent data phrases. All participants experienced all the proposed interaction techniques. We report and discuss how the proposed interactions 1) addressed the issues identified in the formative study, 2) could improve participants’ current authoring workflows, and 3) could be extended for data exploration and to enable new workflows that bridge the gap between the writing and data exploration stages. We also report on observed behaviors that suggested future improvements for real-world usage.

\subsubsection{Utility of Text-Data Connections}
\noindent
The interaction techniques provided by CrossData were lauded and rated as useful by participants (\autoref{fig:study}). 
Participants confirmed that these techniques addressed key pain points in their daily workflows and considered them to be \quot{killer features} (E5) for writing data documents. 
Among the various techniques, participants appreciated the compute value (7/8 strongly agree, 1/8 agree) and retrieve value (6/8 strongly agree, 2/8 agree) techniques as they facilitated the inputting of data (T1) by \quot{enable[ing] computation using words (E8)}, 
\quot{reduc[ing] application switching} (E8), 
and \quot{avoid[ing] typos} (E3). 
As commented by E3, these techniques addressed some \quot{fundamental issues} and thus brought \quot{fundamental improvements to the writing process.}

Participants also responded positively (4/8 strongly agree, 4/8 agree) to the techniques designed to maintain consistency between data and text (T2). 
These techniques helped users \quot{ensure consistency} (E3) with \quot{fewer manual efforts} (E1).
E6 believed that these techniques could help her company \quot{reduce human resource costs on the review team}.

The interactive techniques that facilitated iteration (T3) via interaction with data-driven text (5/8 strongly agree, 3/8 agree) and the automatic adjustments of tables (5/8 strongly agree, 3/8 agree) and charts (5/8 strongly agree, 3/8 agree) were also appreciated by participants because these techniques could \quot{significantly reduce working back-and-forth} (E5) and enabled participants to \quot{rapidly refine the charts [and tables]} (E7).
Participants (E1, E4, E7) also remarked that the interactivity of the text, as well as the real-time synchronization between text, table, and charts, made the authoring process \quot{fun and engaging} (E1), but also could assist in thinking processes and inspire more ideas during writing as the user can \quot{see what he is writing} (E4).

\subsubsection{Authoring Workflow vs Traditional Tools}
\noindent
All participants agreed that the interactions provided by CrossData would mesh well with their current workflows (4/8 strongly agree, 4/8 agree), e.g., \quot{you just need to write as usual} (E1). 
They further commented that these interaction techniques did not require installing another application and could be easily integrated within existing tools by \quot{installing [them as] a plugin to my Word} (E2).

All participants found that the interaction techniques could streamline their workflows due to \quot{less context switching} and allow for efficient iterations of a document. 
E8 noted that she used to frequently switch between \quot{Excel, Word, and sometimes the calculator} during the writing process, which was \quot{stressful and distracting.}
By integrating CrossData with the existing tools, participants could \quot{concentrate on her writing} (E8), and \quot{focus on the current writing without worrying about refining or updating other sentences} (E7).

Another improvement to participants’ workflows that was mentioned was \quot{facilitating the process of getting feedback from others} (E2). 
Mainstream tools such as Word and PowerPoint present reports in a static manner and thus hinder authors from addressing or responding to others’ feedback immediately, whereas the features provided by CrossData \quot{make it very useful to answer ad-hoc questions during the discussions that would normally require some follow up work, e.g., swap out regions, look at percentage changes between different time periods, etc.} (E5)

In terms of the negative impacts these techniques may have on their workflows, E7 noted that \quot{perhaps the only cost is to learn how to use [them]}. Specifically, \quot{you need to understand the concepts and get familiar with, for example, placeholders} (E7). 
Nevertheless, all participants reported that the interaction techniques were easy to learn and easy to use (\autoref{fig:study}), indicating that the downside of using them would be negligible.

\subsubsection{Enabling New Workflows to Bridge the Gap between Data Exploration and Writing}
\noindent
While CrossData was designed to support the writing stage, the intertwined nature of exploration and writing inspired participants to imagine CrossData beyond the presented tasks, and they suggested several benefits that could be enabled by the language-oriented techniques to facilitate data analysis and exploration. 

First, natural language enables one to express reusable high-level goals instead of performing transient low-level operations, thereby improving the efficiency of data exploration. 
E4 noted that with the compute value technique provided by CrossData, he could efficiently calculate a value by typing a sentence instead of \quot{scroll up and down in a sheet and brush and re-brush the cells.} 
Moreover, E3 suggested that the exploration process could be easily reused for different data by copying and pasting the text, i.e., \quot{I can write text to retrieve and calculate values, and then copy the text to another sheet to get new values ... this is impossible in Excel since I cannot copy my interactions on one sheet to another.}

Second, CrossData could facilitate active thinking during the exploration process. 
E1 found that the suggestion list and interactive operators inspired him to explore the data from new angles that he missed before. He remarked that the suggested text was similar to the query recommendations in search engines. 
E4 explained that sometimes he stopped data exploration because it required too many tedious operations with Excel, i.e., \quot{exploration is a process of thinking rather operating the Excel…I will definitely explore more if only a few clicks or types are required.} 

Third, language-oriented data exploration enabled users to \quot{record their exploration process as [a] draft} (E1) and naturally \quot{shift from data exploration to writing.} 
All participants confirmed that there was a gap between data exploration and presentation in their current workflow, which has been recognized in prior work as an important research direction to improve the workflow of data analysts~\cite{Latif2021}. 
E7 commented that these \quot{two interconnected stages [i.e., data exploration and communication,] were usually separated in two disconnected applications.} 
With language-oriented interaction techniques, however, data exploration and data document authoring can be tightly integrated such that \quot{exploring [the data] is drafting [the document] and vice versa.}

\subsubsection{Observed Behaviors}
\label{sssec:behaviors}
\noindent
We observed several interesting user behaviors that reflected participants’ real-world writing practices that were not supported by our current implementation.

First, when the data operations were simple, participants tended to directly type the result, which could result in untracked connections. For example, when writing \quot{The U.S. reports the most new cases in America}, E3 manually typed \quot{The U.S.} instead of using the placeholder feature. This was because that the participant already knew the desired data, and inserting a placeholder required more effort. 
The result, however, was that \quot{The U.S.} text would not be updated when the participant was asked to modify \quot{America} to \quot{Africa}, causing data inconsistency due to the missing connections. While the Connection Engine is currently designed to interactively recommend data phrases, to address this issue, it could be extended to detect and connect manually typed dependent phrases to ensure all data phrases would be connected with the underlying dataset.

We also observed that some participants reported approximate numbers instead of exact data values, which caused undesired suggestions from the engine. For example, E1 wrote that \quot{[Placeholder] countries in America report more than 10,000 ...}. He wanted to connect \quot{10,000} with the new cases column. However, because \quot{10,000} is an approximate number that did not exist in the new cases column, the Connection Engine could not return suggestions because it relies on string and synonym matching to suggest independent phrases. E1 then struggled to connect the \quot{10,000} with the new cases column. Such behavior was also observed in other participants (e.g., E2, E5, and E7). 
While participants altered the approximate numbers to exact values to create connections, this issue could be common in real-world scenarios. To address this, CrossData could be extended to allow users to manually insert their desired connections or support fuzzy data value matching when certain keys are present, such as ``almost'' and ``more than''.

Third, the participants tended to write safe, simple sentences to ensure the connections would be created successfully during writing. Overall, the sentences were relatively simple and had similar structures to the sentences in the training and reproduction tasks. 
While this could be attributed to the limited time frame of the task, it is possible that participants faced a dilemma when guessing which written text the system could understand and establish connections with. 
Such an issue has been recognized as a long-standing challenge for users of NLI systems~\cite{Discoveringnlp}.
To address this issue, we propose that the system could provide alternative methods (e.g., interface actions) to allow users to manually create text-data connections instead of fully relying on the auto-extraction of the connections from the text. 
Several participants confirmed this improvement would be useful and necessary in their interviews, indicating that \quot{the system should enable users to create or modify the connections after the writing.} (E7)

\subsubsection{Limitations}
\noindent
Participants noted several limitations of CrossData and suggested some improvements. Similar to other interactive systems that employ NLP, 
CrossData can misinterpret users’ intentions due to the reasons discussed in Section~\ref{ssec:failure} and Section \ref{sssec:behaviors} (e.g., lack of context, unrecognized approximate numbers).
While CrossData allows users to correct misdetections caused by predefined rules, it does not support the correcting errors caused by NLP techniques. 
All participants expressed their concern regarding this and understood that they could be mitigated by further advancements of NLP techniques, more intelligent connection recognition algorithms, and by them being able to flexibly modify the suggested connections.

Participants also proposed several improvements with regards to extensibility and customizability. For example, E8 suggested that CrossData could support customized operators and calculations or enable users to import domain-specific operators from online libraries. E3 proposed that CrossData should enable users to share their customized operators with others to facilitate collaborative editing. E5 indicated that the system should enable users to "freeze" connections so that they could rephrase sentences without worrying about losing any connections. 

Several participants also raised the concerns about scalability. For instance, E1, an auditor, who often needed to write data documents to synthesize findings from more than 50 datasets, noted that connecting a phrase to all underlying datasets could lead to too many possible connections.
A potential solution to this could be to add a context-awareness mechanism to CrossData so that it could prune the search space based on one's writing context, e.g., the surrounding sentences, tables, charts, and section titles. 

\section{Future Work}

\vspace{-1mm}
\paragraph{\textbf{Beyond Tabular Data and Basic Charts}}
CrossData currently supports the connection of text to tabular data, wherein each data item is represented as a row and its attributes are represented as columns. While tabular data is common in practice, it does not naturally contain information about the rich relationships that exist among data items and are often found within graph-based or tree-based data structures. One future direction for language-oriented authoring research could be to support users in connecting text to rich data structures. The data visualizations currently supported within CrossData are basic charts (e.g., line and bar charts), however, future work should explore how to support more customized, complex data visualizations. This, of course, would require the identification of mappings between the natural human language used in data documents and the domain-specific terms used during data analysis and visualization processes. To develop such mappings, we plan to collect and annotate existing data documents that describe or contain various data structures and visualizations. 

\vspace{-2mm}
\paragraph{\textbf{Blurring the Line between Writing and Programming for Data Analysis and Visualization}}
In addition to graphical user interface applications, programming is another commonly used modality for data analysis and visualization. For example, computational notebook applications, which enable users to write programs to analyze and visualize data, are becoming increasingly popular. A common practice when using computational notebooks is to write explanatory textual descriptions alongside a program’s code to facilitate documentation and collaboration.  This presents an opportunity to extend the use of written text for data analysis and visualization. Thus, one future direction could be to integrate CrossData into computational notebooks, so that users can analyze and visualize data by writing descriptive and self-explanatory text without requiring programming skills.

\vspace{-2mm}
\paragraph{\textbf{Supporting Dynamic and Interactive Data Presentations}}
While CrossData leveraged text-data connections to support the authoring of static data documents, 
the data documents that results were interactive,
suggesting opportunities to create interactive documents without any programming.
We plan to expand CrossData to further support the creation of data-driven diagrams and simulations. Another future direction will be to explore the creation of other forms of dynamic and interactive presentations of data with text-data connections, such as data videos and data animations. Specifically, the connections between text with tables and charts could be directly employed to create animated changes in tables and charts that correspond with the narration of animation, videos, or slideshows.
\section{Conclusion}

Despite the proliferation of applications and systems that seek to support users while analyzing, visualizing, and communicating data, the authoring of data documents still remains a laborious process. Within this work, we conducted a formative study with eight professionals and found that a key reason for these tedious, repetitive, and error-prone workflows is the lack of connections that exist between text and data. As the process of writing data documents is implicitly the process of establishing connections between text and data, we thus developed a prototype system, CrossData, that would infer text-data connections within written text. The development of CrossData enabled for a systematic exploration of the power of identifying, establishing, and reifying text-data connections as persistent, interactive, first-class objects that could be used to assist in the authoring of dynamic, interactive data-driven documents. 
A technical evaluation demonstrated the effectiveness and robustness of the connection engine.
Results from an expert evaluation found that CrossData not only reduced the manual effort required while writing data documents, but also provided new opportunities to leverage text to support the authoring of data-driven content.

\bibliographystyle{ACM-Reference-Format}
\bibliography{cd-bibtex}

\end{document}